\documentclass[10pt]{article}

\usepackage{amsmath}
\usepackage{amssymb}

\usepackage{cite}

\usepackage{hyperref}

\usepackage{lineno}

\usepackage{microtype}
\DisableLigatures[f]{encoding = *, family = * }

\usepackage{graphicx}

\usepackage[labelfont=bf,labelsep=period,justification=raggedright]{caption}

\makeatletter
\renewcommand{\@biblabel}[1]{\quad#1.}
\makeatother

\date{}

\begin{document}

\begin{flushleft}
{\Large
\textbf{Bacterial cartels at steady supply}
}
\\
Thibaud Taillefumier$^{1}$, Anna P\'osfai$^{1}$, Yigal Meir$^{2}$
Ned S. Wingreen$^{1,3,\ast}$
\\
\textbf{1}  Lewis-Sigler Institute for Integrative Genomics, Princeton University, Princeton, NJ 08544, USA
\\
\textbf{2} Department of Physics, Ben-Gurion University of the Negev, Beer Sheva 84105, Israel
\\
\textbf{3} Department of Molecular Biology, Princeton University, Princeton, NJ 08544, USA
\\
$\ast$ Email: \href{mailto:wingreen@princeton.edu}{wingreen@princeton.edu}
\end{flushleft}

\section*{Abstract}

Metagenomics has revealed hundreds of bacterial species in almost all microbiota. 
In a few well-studied cases, bacterial communities have been observed to coordinate their metabolic fluxes. 
In principle, bacteria can divide tasks to reap the benefits of specialization, as in human economies. 
However, the benefits and stability of an economy of bacterial specialists are far from obvious.
Here, we physically model the population dynamics of bacteria that compete for steadily supplied resources. 
Importantly, we explicitly model the metabolic fluxes yielding cellular biomass production under the constraint of a limited enzyme budget.
In our framework, we find that population dynamics generally leads to the coexistence of different metabolic types, which satisfy an extended competitive exclusion principle (even allowing for adaptive mutation). 
We establish that these consortia act as cartels, whereby population dynamics pins down resource concentrations at values for which no other strategy can invade.
Finally, we propose that at steady supply, cartels of competing strategies automatically yield maximum biomass, thereby achieving a collective optimum.

\section*{Significance}

In human economies, cartels are formed to avoid competition by controlling resource availability.
Building on a physical model for resource-limited growth, we show that metabolic competition between bacteria similarly leads to the selection of cartels that control resource availability via population dynamics.
Specifically, cartels avoid competition by pinning down resource concentrations at values for which no metabolic variant can outcompete the cartel's members.
We propose that cartels also yield maximum biomass, constituting a microbial example of Adam Smith's ``invisible hand" leading to collective optimal usage of resources.
Our analysis illustrates how division of labor among distinct metabolic types can be predicted from optimization principles.
These optimization principles, derived from transport-network theory, may provide a guide to understanding the division of labor in complex bacterial communities.


\section*{Introduction}

Bacterial diversity is ubiquitous. 
Every gram of soil or litter of seawater contains hundreds or more microbial species \cite{Daniel:2005aa}. 
In humans, the gut microbiome comprises at least $500$ microbial species \cite{Lozupone:2012aa}.
These diverse microbial communities are widely credited with division of labor, collectively reaping the benefits of specialization by dividing tasks among different organisms.
In a few well-studied cases, bacterial communities have been observed to coordinate their metabolic fluxes \cite{Foster:2011fk}.
For instance, in bacterioplankton communities, heterotrophic species collectively coordinate their metabolism during the day-night cycle \cite{Ottesen:2014}.
Shotgun sequencing has begun to unveil the biochemical networks at work in complex environmentally sampled communities \cite{Tyson:2004ij,Gill:2006}.
However, the lack of knowledge about gene functions and gene distributions in individual cells hinders the interpretation of this data \cite{Cordero:2014bs}.

There are also serious conceptual challenges to understanding diversity in metabolically competing microbial communities.
For instance, the emergence of diversity in ``consumer-resource'' models is limited by the competitive exclusion principle:  at stationary state, the number of coexisting species cannot exceed the number of available resources \cite{Hardin:1960ly, Levin:1970aa}.
This principle severely limits diversity in models that consider a few resources as in the case of the ``paradox of the plankton'' \cite{Petersen:1975}.
Another essential challenge is understanding the persistence of microbial diversity in the face of potentially more fit metabolic variants; these reinforce the challenge posed by the competitive exclusion principle: 
in consumer-resource models, a fitter strain colonizes a niche at the expense of those already present by depleting the pool of essential resources, generally leading to a collapse in diversity \cite{Shoresh:2008ys}.

The above conceptual challenges call for a physically-based model for competing metabolic strategies. 
However, classical consumer-resource models generally prescribe the rate of biomass production via phenomenological functions of the abundances of essential resources without explicit conservation of fluxes \cite{Liebig:1840,Monod:1950aa}.
Here, we introduce a flux-conserving physical model for bacterial biomass production to address two intertwined questions: Can the structure of metabolic networks explain the emergence of bacterial division of labor? And what efficiencies can bacteria achieve via such a division of labor?

Considering that biomass (primarily protein \cite{Simon:1989}) results from the assembly of building blocks (amino acids or amino acid precursors), we explicitly model the  fluxes associated with the metabolic processing of these building blocks, including enzyme-mediated import and conversion \cite{Almaas:2004pi}.
Different metabolic strategies are defined by specific distributions of these enzymes, which collectively satisfy a budget constraint.
We find that at fixed building-block supply, competitive population dynamics leads to the stable emergence of bacterial consortia, with at least as many distinct metabolic strategies as there are shared resources.
Importantly, such consortia form cartels that resist invasion by metabolic variants.
We employ optimization principles from transport-network theory to elucidate the structure of these cartels, relating the stable metabolic strategies to the ordering of external building-block availabilities.
We show that the metabolic strategies employed by cartels collectively control external resource concentrations.
This suggests that one benefit of metabolic diversity stems from the ability of cartels to shape their environment to achieve a stable collective biomass optimum.


\section*{Model}

\begin{figure}
\centerline{\includegraphics[width=.6\textwidth]{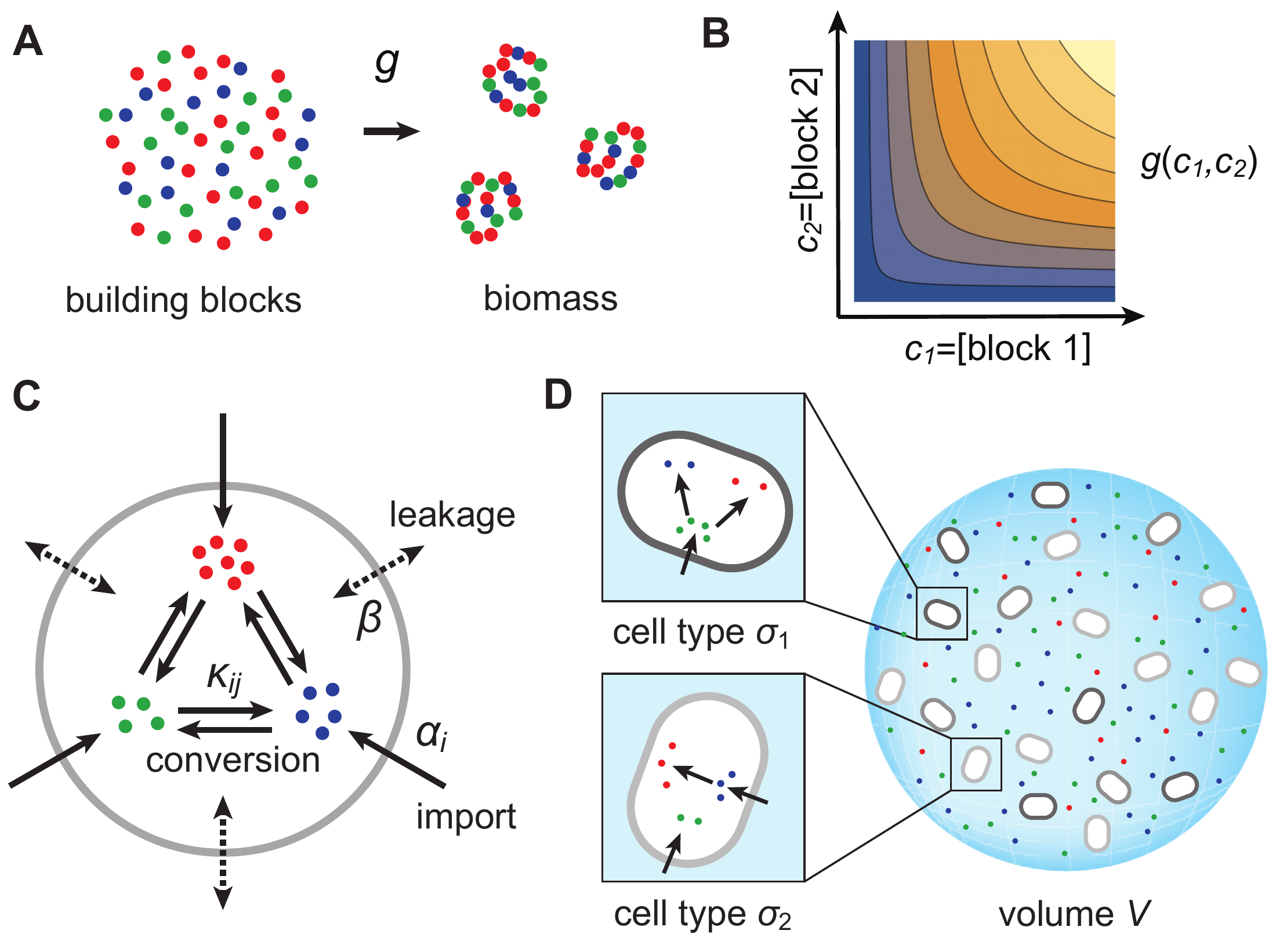}}
\caption{Model for metabolically competing cell types. 
({\it A}) The rate of biomass production $g(c_1, \ldots, c_p)$ is a function of the internal building-block concentrations.
({\it B}) Biologically relevant growth-rate functions $g(c_1, \ldots, c_p)$ are increasing with respect to $c_i$ with diminishing returns.
({\it C}) Different cell types, i.e. metabolic strategies, are defined as specific distributions of enzymes for import  $\alpha_i$ and conversion $\kappa_{ji}$, subject to a finite budget.
({\it D}) Cell types (e.g. $\sigma_1$ and $\sigma_2$) compete for external building blocks that are steadily and homogeneously supplied in volume $V$.
\label{fig:model}
}
\end{figure}

In this section, we present a model for the population dynamics of cell types metabolically competing for external resources (see Fig.~\ref{fig:model}). Importantly, biomass production is governed by a physical model that respects flux conservation.

\subsection*{Resource-limited growth model}

As cellular growth is primarily due to protein biosynthesis, we consider biomass production to result from the incorporation of building blocks (amino acids or amino acid precursors) into biologically functional units (proteins).
Specifically, we assume that biomass production requires $p$ types of building blocks and we denote by $b_i$, $1\leq i \leq p$, the concentration of block $i$ in cellular biomass.
To maintain the stoichiometry of building blocks in biomass, bacteria that grow at rate $g$ incorporate block $i$ at rate $g b_i$.
As the incorporation of building blocks is limited by the internal availability of free building blocks, we model the growth rate as a function $g\left(c_1, \ldots, c_p\right)$, where $c_i$ is the internal concentration of block $i$.
To obtain a plausible functional form for $g\left(c_1, \ldots, c_p\right)$, we consider the rate of incorporation of a building block to be proportional to its concentration.
Then the time to produce a unit of biomass (e.g. a protein) is the sum of the incorporation times for each type of block $i$, which we take to be proportional to $b_i/c_i$, the ratio of the building-block concentration in cellular biomass to the internal free building-block concentration. 
The growth rate, which is proportional to the inverse of this time, therefore has the form 
\begin{eqnarray}\label{eq:harm}
g(c_1, \ldots,c_p) = \gamma \left(\frac{b_1}{c_1}+ \ldots + \frac{b_p}{c_p}\right)^{-1} \, ,
\end{eqnarray}
where $\gamma$ is a rate constant.
For simplicity, we consider that all bacteria use the same molecular machinery and building-block stoichiometry to produce biomass.
Thus, we consider that the rate function $g(c_1, \ldots,c_p)$ is universal, independent of the metabolic strategy used by a bacterium to accumulate building blocks.

In order to accumulate block $i$ internally, a bacterium can import block $i$ from the external medium or produce it internally via conversion of another building block $j$. Thus, to produce biomass, bacteria can substitute resources for one another. 
We allow all possible imports and conversions.
The quantitative ability of a bacterium to import and convert building blocks constitutes its ``metabolic strategy'', and corresponds to the cell's expression of the enzymes that mediate building-block import and conversion. 
For simplicity, we assume that metabolic fluxes are linear in both enzyme and substrate concentrations.
This assumption  corresponds to enzymes operating far from saturation, which is justified during resource-limited growth.
Specifically, denoting the external concentration of block $i$ by $c_i^{\mathrm{ext}}$, the enzyme-mediated fluxes associated with import and conversion of block $i$ have the form $\alpha_i c_i^{\mathrm{ext}}$ and $\kappa_{ji} c_i$, respectively, where $\alpha_i$ and $\kappa_{ji}$ are enzymatic activities, which are proportional to the number of enzymes dedicated to each metabolic process.
In addition to enzyme-controlled fluxes, we include passive leakage across cell membranes \cite{Paczia:2012tg,Reaves:2013fk}, with net influx $\beta (c_i^{\mathrm{ext}}-c_i)$.
As cells can only devote a certain fraction of their resources to the production of enzymes, we require the enzymatic activities of each bacterium to satisfy a budget constraint, 
\begin{eqnarray}\label{eq:budCons}
\sum_i \alpha_i + \sum_{j,i} \kappa_{ji} \leq E \, ,
\end{eqnarray}
where $E$ denotes the total enzyme budget.
The metabolic strategy of a cell type $\sigma$ is specified by a set of enzyme activities $\lbrace \alpha_{i,\sigma}, \kappa_{ij,\sigma} \rbrace$ that satisfy the budget constraint \eqref{eq:budCons}.

\subsection*{Conservation of building blocks}

We consider various cell types $\sigma$ growing in an homogeneous environment of volume $V$.
We denote the dimensionless population count of cell type $\sigma$ by $n_\sigma$ and the total population count by $N = \sum_\sigma n_\sigma$. 
In the volume $V$, we consider that the $p$ building blocks are steadily supplied at rates $s_i$ (concentration/time) and can be lost, e.g. via degradation and/or diffusion out of the volume, at a rate $\mu$. 
Each cell type processes building blocks according to its own metabolic strategy.
Conservation of internal building block $i$ for cell type $\sigma$ prescribes the dynamics of the internal concentration $c_{i,\sigma}$ (see Supporting Information),
\begin{eqnarray}\label{eq:intFlux}
\dot{c}_{i,\sigma} = (\alpha_i+\beta) c^{\mathrm{ext}}_i-\beta c_{i,\sigma} - \sum_{j \neq i} \kappa_{ji} c_{i,\sigma} + \sum_{j \neq i} \kappa_{ij} c_{j,\sigma} - g b_i \, ,
\end{eqnarray}
where the only nonlinearity is due to the growth function $g$.
Populations of the various cell types exchange building blocks with the external resource pool via import and leakage, and also via biomass release upon cell death \cite{Simpson:2007aa,Schutze:2013aa}.
Conservation of extracellular building block $i$ prescribes the dynamics of the external concentration $c^{\mathrm{ext}}_i$ (see Supporting Information),
\begin{eqnarray}\label{eq:extFlux}
\dot{c}^{\mathrm{ext}}_i = s_i -  \mu  c_i^{\mathrm{ext}} - \frac{v}{V-Nv}\left(\sum_\sigma n_\sigma\phi_{i,\sigma} \right)  \, ,
\end{eqnarray}
with cell-type-specific fluxes 
\begin{eqnarray}\label{eq:indFlux}
\phi_{i,\sigma}=  (\alpha_{i,\sigma} + \beta) c_i^{\mathrm{ext}}- \beta  c_{i,{\sigma}} - f \delta b_i \, ,
\end{eqnarray}
where $\delta$ is the rate of cell death (assumed constant) and $f$ is the fraction of biomass released upon cell death.
Per-cell fluxes $\phi_{i,\sigma}$ contribute to changing the external concentration $c_i^{\mathrm{ext}}$ via a geometric factor $v / (V-Nv)$, the ratio of the average individual cellular volume $v$ to the total extracellular volume $V-Nv$.
As intuition suggests, the smaller the number of cells of a particular type, the less that cell type impacts the shared external concentration via metabolic exchanges.

\subsection*{Competitive population dynamics}

The inverse of the cellular death rate $\delta$, i.e. the lifetime of a cell, is much larger than the timescales associated with metabolic processes such as building-block-diffusion, conversion, and passive/active transport.
This separation of timescales justifies a steady-state approximation for the fast-variables: $\dot{c}_{i,\sigma}=0$ and $\dot{c}_i^\mathrm{ext}=0$.
With this approximation, Eqs. \eqref{eq:intFlux} and \eqref{eq:extFlux} become flux-balance equations for building blocks.
Solving Eq. \eqref{eq:intFlux} with $\dot{c}_{i,\sigma}=0$ yields the internal concentrations $c_{i,\sigma}(c_1^{\mathrm{ext}}, \ldots, c_p^{\mathrm{ext}})$ as increasing functions of the external concentrations.
In turn, solving Eq. \eqref{eq:extFlux} with $\dot{c}_{i,\sigma}=0$ and using the functions $c_{i,\sigma}(c_1^{\mathrm{ext}}, \ldots, c_p^{\mathrm{ext}})$ yields the external concentrations $c_i^{\mathrm{ext}}(\lbrace n_\sigma \rbrace)$, as well as the growth rates of cell types $g_\sigma(\lbrace n_\sigma \rbrace)$, as functions of the populations of cell types.
Hence, the population dynamics of the cell types is described by a system of ordinary differential equations 
\begin{eqnarray}\label{eq:popDyn}
\frac{\dot{n}_\sigma}{n_\sigma}
=
g_\sigma \big( \lbrace n_\sigma \rbrace \big) - \delta \, 
\end{eqnarray}
that are coupled via the external concentrations.
Note that the population dynamics is driven and dissipative: building blocks are constantly both supplied to and lost from the external media, while cell death leads to loss of building blocks because only a fraction of biomass is recycled $(f<1)$. 



\section*{Results}

In this section, we show that competitive population dynamics generically select for bacterial consortia with distinct metabolic types, and we characterize the structure and benefit of the division of labor in these consortia.


\subsection*{Numerical simulations}

We investigated the possibility of stable coexistence at steady supply rates by simulating competitive population dynamics subject to invasion by new metabolic variants.
In our simulations, cell types have distinct metabolic strategies defined by randomly chosen enzyme distributions $\lbrace \alpha_i$, $\kappa_{ji} \rbrace$ satisfying the budget constraint \eqref{eq:budCons}, with universal growth-rate function \eqref{eq:harm}.
If the number of distinct strategies exceeds the number of resources then some cell types become extinct: $n_\sigma<1$.
Whenever a cell type $\sigma$ is driven to extinction, we replace it with another randomly sampled strategy $\sigma'$ with $n_\sigma'=1$, thereby modeling invasion by metabolic variants.

\begin{figure}
\vspace{-5pt}
\centerline{\includegraphics[width=.6\textwidth]{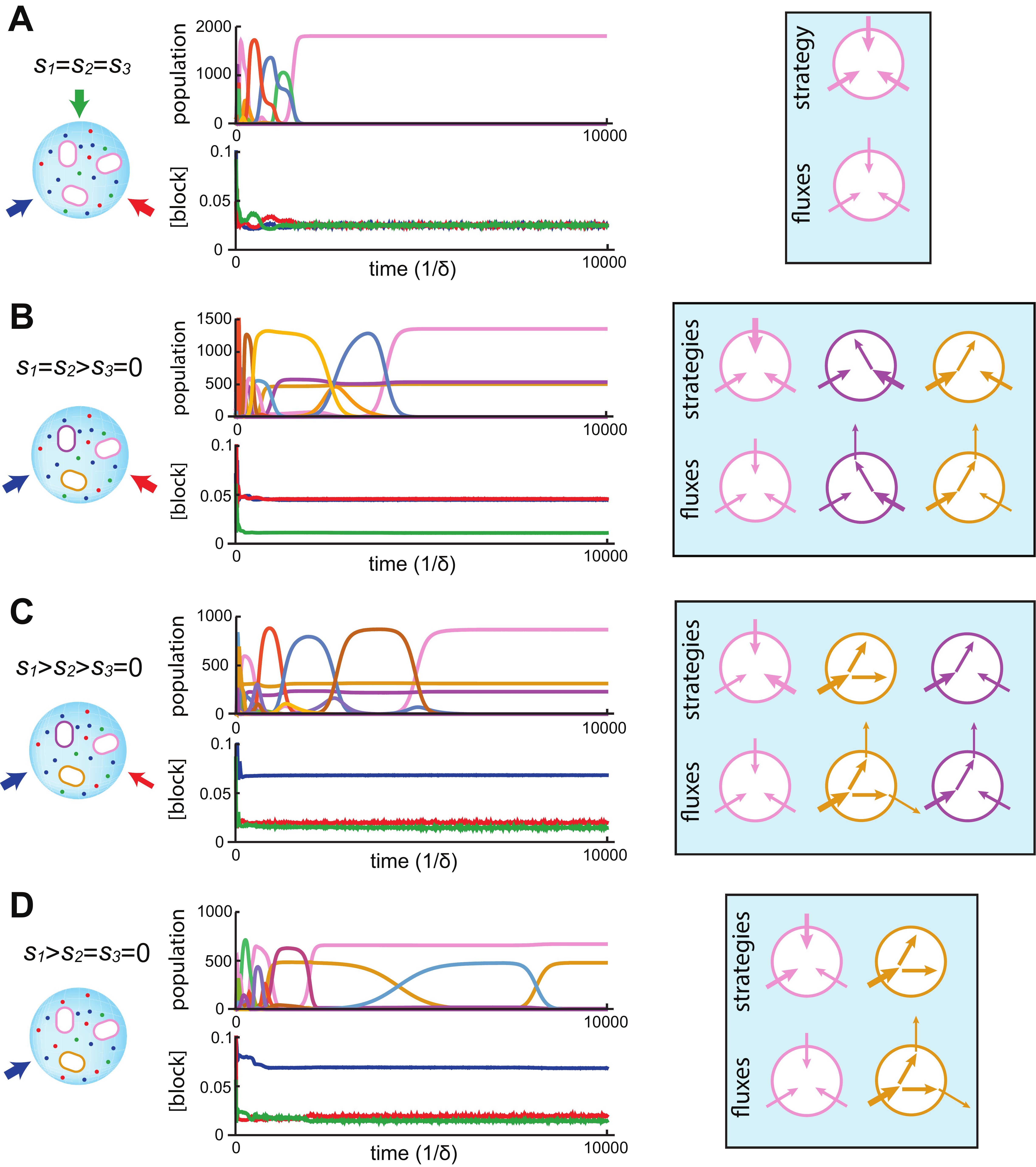}}
\caption{Simulated competitive dynamics.
In all panels, the left schematic indicates supply conditions, the central plot shows an example of competitive population dynamics, and the right diagram depicts the strategies and their internal building-block per-cell fluxes. 
({\it A})  If building blocks are supplied with equal stoichiometry $s_1=s_2=s_3$, metabolic strategies that carry out conversions are wasteful and a single pure-importer cell type prevails.
 ({\it B}) If two building blocks are supplied with equal stoichiometry, e.g. $s_1=s_2>0$ and $s_3=0$, three cell types can coexist: two ``symmetric'' types using supplied blocks as a precursor for block $3$, which accumulates externally due to leakage and release upon cell death, 
and, if $c^{\mathrm{ext}}_3$ is large enough, a third pure-importer type.
({\it C}) For large enough imbalance in the supply of building blocks $1$ and $2$, e.g. $s_1>s_2>s_3=0$, three distinct cell types can coexist:
a pure-converter type imports block $1$ and converts blocks $2$ and $3$; if $c^{\mathrm{ext}}_2$ is large enough, a mixed type emerges, importing blocks $1$ and $2$, and converting $1$ to $3$; and, if $c^{\mathrm{ext}}_3$ is large enough, a pure-importer type.
({\it D}) If only one building block is supplied, e.g. $s_1>s_2=s_3=0$, two strategies coexists:
a pure-converter type releases blocks $2$ and $3$, which can lead to the emergence of a pure-importer type.
The external building-block concentrations fluctuate due to the invasion by and extinction of metabolic variants.
\label{fig:sim}
}
\end{figure}

Fig.~\ref{fig:sim} shows simulated competitive dynamics for $p=3$ building blocks with different rates of building-block supply.
These simulations reveal that competitive population dynamics can lead to the emergence of bacterial consortia, defined as coexisting cell types that are stable against invasion.
Moreover, the metabolic strategies forming these consortia exhibit network structures that are directly related to external building-block availability.
In the following, we elucidate the emergence and persistence of these consortia at steady state, focusing on consortia with at least $p$ distinct cell types that can control external building-block concentrations, and which we therefore refer to as bacterial ``cartels''.

\begin{figure}
\vspace{-15pt}
\centerline{\includegraphics[width=.6\textwidth]{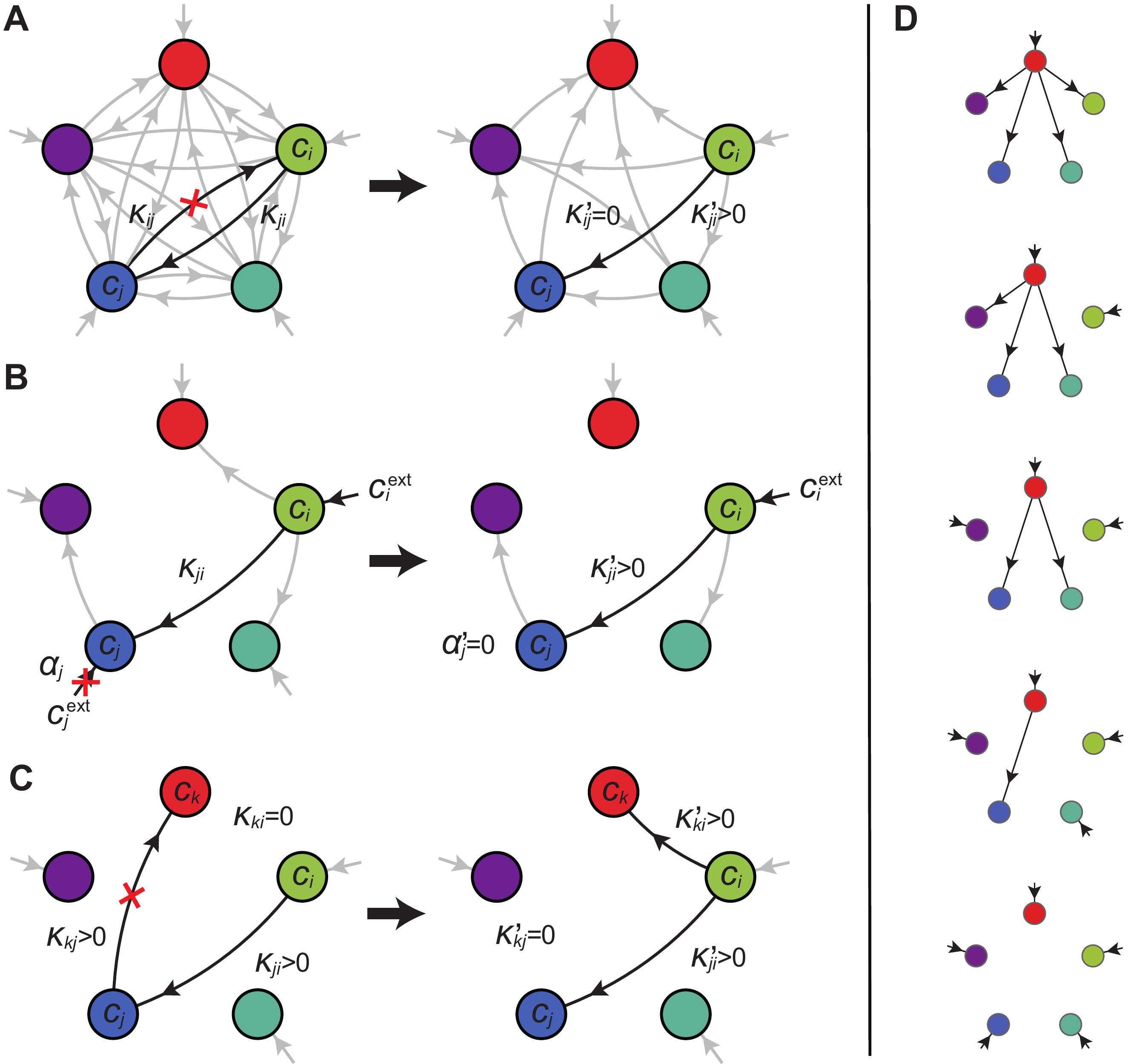}}
\caption{Optimal metabolic classes. 
A metabolic class is defined by the set of enzymes for which $\alpha_i>0$ and $\kappa_{ji}>0$.
If a metabolic class is optimal, i.e. achieves the fastest growth rate, no other metabolic class can achieve the same growth rate with a lower enzyme budget.
({\it A}) 
Optimal metabolic classes cannot have topological $2$-cycles. 
If cell type $\sigma$ (left) is such that the net conversion flux from block $i$ to block $j$ is positive, i.e. $\kappa_{ji} c_i>\kappa_{ij} c_j$, a cell type $\sigma'$ (right) that only differs from $\sigma$ by $\kappa'_{ij}=0$ and $\kappa'_{ji}=\kappa_{ji}-\kappa_{ij} c_j/c_i$ achieves the same growth rate as $\sigma$ but more economically.
More generally, optimal metabolic classes have no topological cycles, i.e. the graphs of their metabolic networks have a tree structure.
({\it B}) Optimal metabolic classes use a single precursor for each converted building block.
If cell type $\sigma$ (left) accumulates block $j$ by import and by conversion from $i$, there is always a more economical strategy $\sigma'$ (right) for which either $\alpha_j=0$ or $\kappa_{ji}=0$.
({\it C}) Optimal metabolic classes convert building blocks in the minimum number of steps. 
If cell type $\sigma$ (left) accumulates block $k$ via a $2$-step conversion from block $i$, there is always a more economical strategy $\sigma'$ (right) that converts block $i$ directly into block $k$.
({\it D}) Optimal metabolic classes can only have a single tree of direct conversion(s).
\label{fig:opt}
}
\end{figure}

\subsection*{Optimal metabolic classes} 

Our simulations of competitive population dynamics suggest that, at steady state, the cell types that form bacterial consortia belong to very specific metabolic classes, as defined by the set of enzymes with $\alpha_i>0$ and $\kappa_{ji}>0$.
In particular, these metabolic classes appear to utilize only a few, non-redundant metabolic processes (many $\alpha_i$ and $\kappa_{ji}$ are zero). 
What is the network structure of these persistent metabolic classes?
Exploiting the linearity of metabolic fluxes, we adapt arguments from transport-network theory \cite{Bohn:2007} to answer this question for an arbitrary number of building blocks (see Fig.~\ref{fig:opt} and Supporting Information).
Specifically, we show that optimal metabolic networks process building blocks via non-overlapping trees of conversions, each tree originating from an imported building-block, and each converted building-block being obtained via the minimum number of conversions.
Intuitively, these properties ensure the minimization of waste (loss of building blocks via leakage) during metabolic processing. 
Moreover, we show that optimal networks use a single building-block resource as precursor for conversions, i.e. there is at most one tree of conversions.
Thus, at steady state, requiring that a metabolic class is optimal, i.e. contains the fastest growing cell type, strongly constrains the graph of its metabolic network.
These constrained graphs can be fully characterized and enumerated for $p$ building blocks:
there are $p$ distinct graphs, each utilizing $p$ distinct enzymes, which defines a total of $1+p(2^{p-1}-1)$ metabolic classes after considering building-block permutations.

\subsection*{Structure of consortia}

To find the composition of bacterial consortia, we must identify the enzyme distributions  $\lbrace \alpha_i$, $\kappa_{ji} \rbrace$  within a metabolic class that yield the fastest growth  for fixed external building-block concentrations.
Obtaining analytical expressions for these optimal enzyme distributions proves intractable for nonlinear growth-rate function such as \eqref{eq:harm}.
However, optimal distributions can be obtained analytically for the minimum model $g(c_1, \ldots, c_p) = \gamma \min(c_1, \ldots, c_p)$, which is closely related to \eqref{eq:harm} (see Supporting Information). 
Knowing analytically the optimal enzyme distributions in each metabolic class allows us to characterize the structure of consortia via the maximum growth rate as a function of external building-block concentrations,
\begin{eqnarray}
G( c_1^{\mathrm{ext}}, \ldots ,c_p^{\mathrm{ext}}) = \max_\sigma g_\sigma (c_1^{\mathrm{ext}}, \ldots ,c_p^{\mathrm{ext}}) \, .
\end{eqnarray}
At competitive stationary state, the maximum growth rate must equal the death rate $\delta$ by Eq. \eqref{eq:popDyn}.
Thus, solving $G( c_1^{\mathrm{ext}}, \ldots ,c_p^{\mathrm{ext}}) = \delta$ determines the set of steady-state external concentrations $c^\star_1, \ldots, c^\star_p$ for which a non-invadable strategy is present.
A non-invadable strategy $\sigma^\star$ is one for which $ g_{\sigma^\star} (c^\star_1, \ldots, c^\star_p) = \delta$.
Consortia emerge for external building-block concentrations for which there is more than one such strategy.

\begin{figure}
\vspace{-20pt}
\centerline{\includegraphics[width=.6\textwidth]{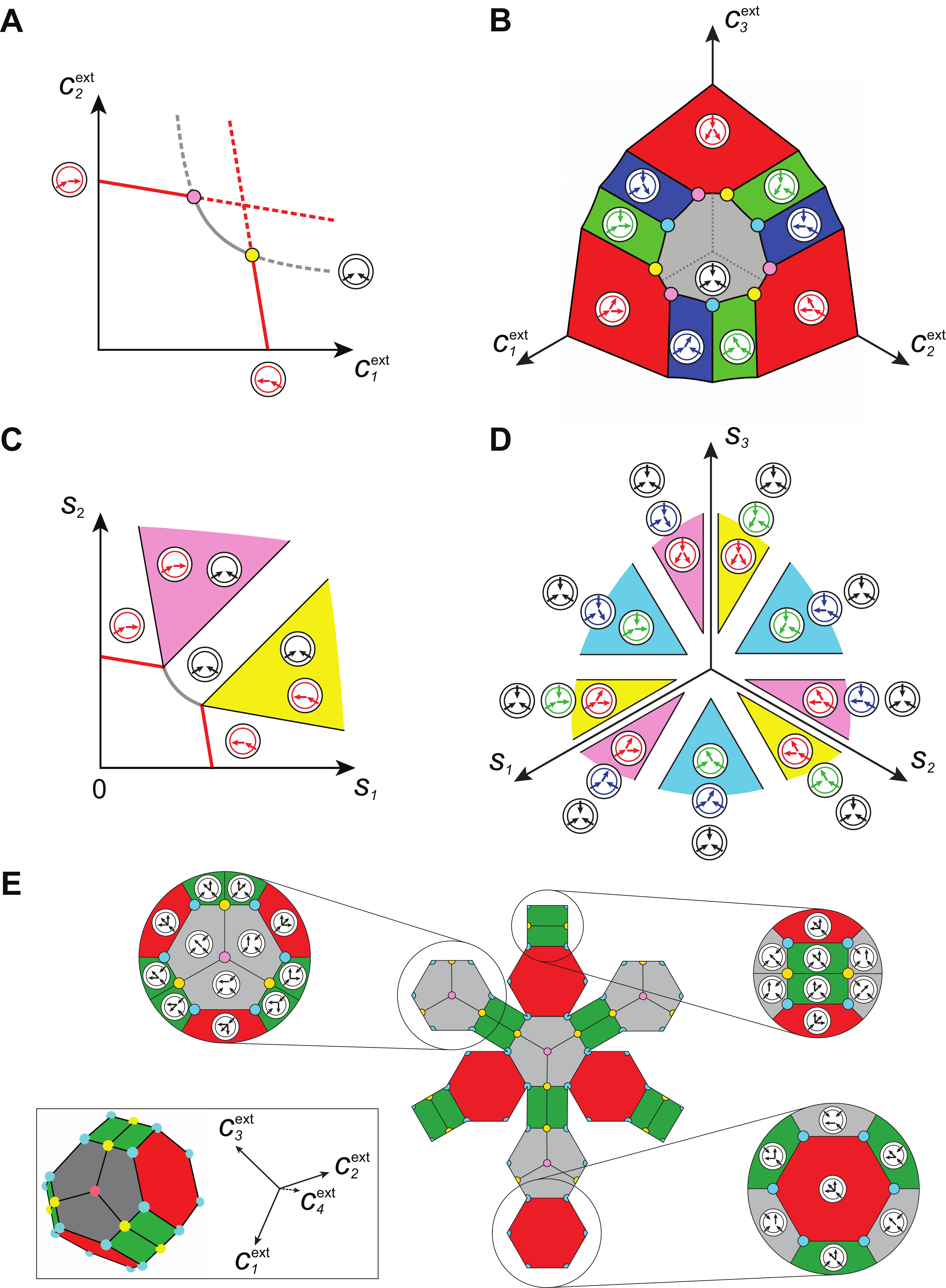}}
\caption{Emergence of bacterial cartels at steady state. 
For large enough supply rates, population dynamics drive the external building-block concentrations towards steady-state values $c^\star_1, \ldots, c^\star_p$ that satisfy growth rate equals death rate, $G (c^\star_1, \ldots, c^\star_p) = \delta$.
Consortia emerge at concentrations for which distinct metabolic classes are jointly optimal.
Cartels are consortia with at least $p$ distinct metabolic classes.
({\it A}) For $p=2$, a pure-converter strategy is optimal on each of the red curves, while a pure-importer strategy is optimal on the grey curve.
Cartels with two distinct cell types exist at the intersection of the grey curve and a red curve. 
({\it B}) For $p=3$, a pure-converter strategy is optimal on the red patches, mixed strategies are optimal on the blue and green patches, while a pure-importer strategy is optimal on the grey patch.
There are two types of cartels at the intersection of $3$ patches: $6$ distinct cartels with well-ordered external concentrations (yellow and pink), e.g. $c_1^{\mathrm{ext}}> c_2^{\mathrm{ext}} > c_3^{\mathrm{ext}}$, and $3$ distinct cartels with degenerate external concentration ordering (cyan), e.g. $c_1^{\mathrm{ext}} = c_2^{\mathrm{ext}} > c_3^{\mathrm{ext}}$.
({\it C})  and ({\it D}): Supply conditions compatible with the emergence of a cartel  for ({\it C}) $p=2$  and ({\it D}) $p=3$.
The set of supply rates for which cartels can arise define non-overlapping polyhedral $p$-dimensional cones, with parallel faces between neighboring cartels, i.e. cartels that share $p-1$ metabolic classes.
Outside of these cones, only fewer than $p$ strategies can survive.
({\it E}) Graph structure of bacterial cartels for $p=4$ building blocks.
As cartels can be labelled by ordering of resource availability, their graph structure is closely related to permutohedron solids, such as the truncated octahedron for $p=4$ (inset: the interior of the truncated octahedron for $p=4$ corresponds to the grey patch shown in ({\it B}) for $p=3$)
In addition to the metabolic types shown, each cartel includes a pure-importer strategy, so that blue and pink cartels have $4$ distinct types while yellow cartels have $5$ distinct types.
In all panels, the circular arrow diagrams depict the metabolic strategies present.
\label{fig:coex}
}
\end{figure}

In Fig.~\ref{fig:coex}, we represent the set of external building-block concentrations compatible with steady states, together with the associated optimal metabolic classes.
For $p$ building blocks (see Supporting Information), we find that there exist $p!$ bacterial cartels, each with $p$ distinct cell types for well-ordered external concentrations, e.g. $c_1^{\mathrm{ext}} > c_2^{\mathrm{ext}} >\ldots > c_p^{\mathrm{ext}}$.
In such cartels, cell type $1$ converts building block $1$ into the $p-1$ other building blocks, cell type $2$ converts building block $1$ into the $p-2$ least abundant building blocks and imports building block $2$, and so forth, and cell type $p$ has a pure-importer strategy.
We also find that for degenerate ordering with $q-1$ equalities, e.g. $c_1^{\mathrm{ext}} = \ldots = c_q^{\mathrm{ext}} > c_{q-1}^{\mathrm{ext}}>\ldots > c_p^{\mathrm{ext}}$, there exist $(p-q)!C^q_p$ bacterial cartels with $1+q(p-q)$ distinct cell types.
In such cartels, cell type $q'$, $1 \leq q' \leq q$ imports all blocks $1, \ldots, q$ but only uses block $q'$ as a precursor for blocks $j>q$, cell type $q''$, $q < q'' \leq 2q$ imports all blocks $1, \ldots, q+1$ but only uses block $q''-q+1$ as a precursor for blocks $j>q+1$, and so forth, and cell type $1+q(p-q)$ has a pure-importer strategy.
Moreover, we find that cartels that share $p-1$ metabolic classes are joined by continuous paths in the space of external concentrations over which these $p-1$ shared metabolic classes remain jointly optimal.
Such paths define a graph which characterizes the topological structure of cartels in relation to changes in external building-block concentrations (see Supporting Information).
Importantly, our analysis shows that the above cartels emerge with the same graph structure for all growth-rate functions satisfying $g(c_1,\ldots,c_p) \geq \gamma \min(c_1,\ldots,c_p)$ for some $\gamma>0$ and having diminishing returns (quasi-concave property), which includes \eqref{eq:harm}.

\subsection*{Relevance of cartels}

Bacterial cartels only exist for specific external building-block concentrations ({\it cf}. the intersection points in Fig. \ref{fig:coex} {\it A} and {\it B}).
\emph{Can competitive population dynamics lead to these cartels for generic supply conditions}?
To answer this question, we compute the set of supply conditions compatible with the emergence of a cartel.
We label a bacterial cartel $\Sigma^\star$ by its associated external concentrations $c_1^\star,\ldots, c_p^\star$, which satisfy a specific (possibly degenerate) order relation.
At concentrations $c_1^\star,\ldots, c_p^\star$, cartel cell types $\sigma \in \Sigma^\star$ jointly achieve the optimal growth rate and are therefore the only surviving cell types.
The per-cell fluxes experienced by these cell types $\phi_{i,\sigma}^\star$ take fixed values that can be obtained via Eq. \eqref{eq:indFlux}.
Then, the resulting flux-balance equations for extracellular building blocks,
\begin{eqnarray}\label{eq:sup}
 s_i \big( \lbrace n_\sigma \rbrace \big)= \mu  c_i^\star + \frac{v}{V-N v} \sum_{\sigma \in \Sigma^\star} n_{\sigma}  \phi_{i,\sigma}^\star  \, ,
 \end{eqnarray}
yield the supply rates as a function of the populations $n_\sigma>0$, $\sigma \in \Sigma^\star$.
In fact, Eq. \eqref{eq:sup} defines the sector of supply rates compatible with the existence of the cartel $\Sigma^\star$ as a $p$-dimensional cone.
Crucially, although $c_1^\star,\ldots, c_p^\star$ specify isolated points in the space of external concentrations, the sectors have finite measure in the space of supply rates, showing that cartels can arise for generic conditions.

In Fig.~\ref{fig:coex}, we represent  the supply sectors associated with each cartel.
For all values of $p$, we find that supply sectors associated with cartels define non-overlapping cones (see Supporting Information).
Moreover cones associated with two connected cartels, i.e. cartels that share at least $p-1$ metabolic classes, have parallel facets in the limit of large budget $E \gg \beta$.
As a consequence, at fixed overall rate of building-block supply $s=s_1+ \ldots +s_p$, the fraction of supply conditions for which no cartel arises becomes negligible with increasing overall supply rate $s$.
For instance, for large rate $s$, every building block has to be supplied at exactly the same rate for a single pure-importer strategy to dominate rather than a cartel.
Therefore, for very generic conditions, a cartel will arise and drive the external building-block concentrations toward cartel-specific values, thereby
precluding invasion by any other metabolic strategy.
This ability to eliminate competition by controlling external resource availability is reminiscent of the role of cartels in human economies, motivating the name ``cartels'' for stable bacterial consortia that include at least $p$ distinct metabolic strategies.

At supply conditions for which no cartel arises, a cell type or a consortium of cell types dominates at steady state but these cell types cannot control external resource availability.
Indeed, a consortium that is not a cartel cannot compensate for changes in supply conditions via population dynamics.
In particular, simply multiplicatively increasing the building-block supply augments the steady-state biomass of a consortium but also modifies steady-state resource availabilities, and therefore the distributions of enzymes that optimally exploit these resources.
In other words, for consortia that are not cartels, optimal metabolic strategies must be fine-tuned to specific supply conditions.

By contrast, within a cartel supply-sector, any increase of the building-block supply is entirely directed toward the cartel's growth of biomass. Remarkably, it appears that bacterial cartels automatically achieve maximum carrying capacity, i.e. they optimally exploit the resource supply. More precisely, for $p=2$ and $3$ building blocks, we found that in each cartel sector, no consortia can yield a larger steady-state biomass than the supply-specific cartel. This result generalizes to arbitrary $p$ if we conjecture that $(i)$ in a cartel sector, the cartelÕs metabolic class can invade any other consortium and that $(ii)$ the optimal growth rates of metabolic class have diminishing returns (see Supporting Information). Intuitively, conjecture $(i)$ means that the emergence of a cartel does not depend on the history of appearance of distinct metabolic classes and conjecture $(ii)$ means that beating diminishing returns requires a switch of metabolic classes. Together, these conjectures ensure that adding a new metabolic calss when possible implies a decrease in the total abundance of building blocks, i.e. a better use of resources. Because a better use of resources is equivalent to a steady-state biomass increase (by virtue of building-block conservation), this establishes that competing bacteria achieve the global collective optimum by forming cartels.



\section*{Discussion}

Building on a physical model for metabolic fluxes, which importantly includes a finite enzyme budget, we showed that competitive population dynamics leads to the emergence of bacterial cartels.
Cartels are defined  as stable consortia of at least as many distinct cell types---each with a fixed metabolic strategy---as there are shared resources.
Within this framework, the benefit of metabolic diversity to the participating cells stems from the ability of cartels to shape the external availability of elementary biomass constituents \cite{Sanchez:2013hc}. 
In particular, cartels can adjust their populations to compensate for changes in supply.
Strikingly, our results support the conclusion that cartels of competing bacteria achieve the optimal collective carrying capacity, as if led by an ``invisible hand'' to efficiently exploit resources \cite{Smith:1776}.

\subsection*{Assumptions and scope of the model}

For simplicity, we assumed linear metabolic fluxes and uniform enzymatic rates, production costs, and building-block stoichiometries.
However, the emergence of optimal cartels does not rely on these assumptions.
Even allowing for fluxes that are nonlinear (e.g. Michaelis-Menten) with respect to building-block concentrations, the enzymes in the pathways that are ultimately rate limiting must operate in the linear regime to be metabolically optimal:
Because resources are depleted by competitive growth, rate-limiting nutrients will fall below saturating levels.
Cells can improve their growth rate by reallocating their enzyme budget from saturated enzymes to the unsaturated enzymes mediating growth-limiting linear fluxes.
Moreover, independent of rates, production costs, and stoichiometries, optimal metabolic types must consist of non-overlapping trees of conversions.
Indeed, the optimality of such metabolic networks, obtained from transport-network theory, only requires the linearity of metabolic fluxes with respect to enzyme concentrations.
As a result, optimal metabolic types, as well as cartels, can still be enumerated.
Interestingly, we discovered that distinct cartels can arise for very similar external building-block availabilities, and cartels can even merge under special conditions.
In an extended model that includes fluctuations, e.g. in enzyme expression \cite{Wang:2011,Kiviet:2014aa}, we expect ``ghosts'' of these neighboring cartels associated with similar resource availabilities to persist against the background of the dominant cartel.

\subsection*{Realistic metabolic networks} 

What relevance might our results have for real metabolic networks?
Bacteria regulate metabolic processes via complex networks with, e.g., multistep reaction chains and metabolic branch points \cite{Almaas:2004pi}.
However, there is evidence of optimal partitioning of enzymes in these real networks: 
bacteria produce components of multiprotein complexes in precise proportion to their stoichiometry, whereas they produce components of functional modules differentially according to their hierarchical role \cite{Li:2014aa}.
Recent experimental studies have revealed that optimal metabolic flux partitioning is an operating principle for resource allocation in the proteome economy of the cell \cite{Hui:2015,Hermsen:2015}.
Provided optimality considerations apply to real metabolic networks, the approach we have taken can provide insight into flux partitioning and division of labor in bacterial communities. 
For instance, we expect that for a group of interconvertible resources that are collectively growth limiting, the expressed metabolic network should have the topological properties discussed above---no cycles, no convergence.
Such predictions can be  tested experimentally by measuring reaction fluxes in large metabolic networks, e.g. using isotope tracers and mass spectrometry.




\subsection*{Spatial and temporal heterogeneities}

Abiotic and biotic processes controlling resource turnover in nutrient reservoirs, such as the ocean or soil sediments, operate on many different temporal and spatial scales \cite{Braswell:1997,Whitman:1998}.
In our framework, steady but spatially inhomogeneous resource supply should lead to the tiling of space by locally dominant cartels.
Because of our model cells' ability to shape their environment, we expect sharp transitions between neighboring tiles, consisting of cartels that differ by a single metabolic class. 
We expect spatial tiling to emerge in real bacterial communities growing in inhomogeneous conditions, e.g. in a gradostat with spatially structured nutrient supply \cite{Lovitt:1981}. 
In such spatial communities, the detection of well-delimited patches of resource availabilities, with specific nutrient ratios, would be evidence of spatial tiling by bacterial cartels.

Temporally varying supply can also be addressed within our framework. 
For supply fluctuations on long timescales $\gg 1/\delta$ (the lifetime of a cell), the population dynamics within cartels keeps resource levels fixed, whereas fluctuations on short timescales  $\ll 1/\delta$ are self-averaging. 
In practice, slow supply fluctuations can arise due to seasonal biogeochemical cycles \cite{Schoener:2011kx}, while fast supply fluctuations can arise from transient biomass release upon cell death \cite{Yoshida:2003aa}.
The effect of supply fluctuations occurring on timescales  $\sim 1/\delta$, which includes day-night cycles, is more complex.
Transport-network theory predicts that fluctuating resource conditions select for networks with metabolic cycles, whose structures depend on the statistics of the driving fluctuations \cite{Katifori:2010,Corson:2010}.
Characterizing the benefit of cycles in such networks may well reveal new optimization principles to better understand bacterial metabolic diversity.


Bacteria also adjust to fluctuating conditions by switching their metabolic type via gene regulation instead of relying on population dynamics.
Within our framework, to consistently implement the optimal mix of metabolic strategies, the role of sensing and regulation is then primarily to determine the relevant ``supply sector'' by assessing the relative abundance of various resources.
The successful completion of this program presumably require coordination via cell-to-cell communication.
We therefore anticipate that extension of our analysis to fluctuating supply conditions may provide insight into the design principles underlying regulation and signaling in bacterial communities.

\section*{Acknowledgments}

This work was supported 
by the DARPA Biochronicity program under Grant D12AP00025, 
by the National Institutes of Health under Grant R01 GM082938,
and by the National Science Foundation under Grant NSF PHY11-25915.
We thank Bonnie Bassler, William Bialek, Curt Callan, and Simon Levin for many insightful discussions.


\end{document}